\documentclass[12pt]{iopart}
\usepackage{amssymb}
\usepackage{graphicx}
\usepackage{epsfig}
\usepackage{color}

\begin{document}

\title{Application of the complex scaling method in solving three-body Coulomb scattering problem}
\author{R. Lazauskas $^{1}$}

\begin{abstract}
The three-body  scattering problem in Coulombic systems is
widespread, however yet unresolved problem by the mathematically
rigorous methods. In this work this long term challenge has been
undertaken by combining distorted waves and Faddeev-Merkuriev
equation formalisms in conjunction with the complex scaling
technique to overcome difficulties related with the boundary
conditions. Unlike the common belief, it is demonstrated that the
smooth complex scaling method can be applied to solve three-body
Coulomb scattering problem in a wide energy region, including
fully elastic domain and extending to the energies well beyond
atom ionization threshold.

Newly developed method is used to study electron scattering on
ground states of Hydrogen and Positronium atoms as well as a
$e^+$+H(n=1) $\rightleftarrows$ p+Ps(n=1) reaction. Where
available, obtained results are compared with the experimental
data and theoretical predictions, proving accuracy and efficiency
of the newly developed method.
\end{abstract}

\address{$^{1}$ Universit{\'e} de Strasbourg, CNRS, IPHC UMR 7178, F-67000
Strasbourg, France \\} \ead{rimantas.lazauskas@iphc.cnrs.fr} \vspace{10pt}
\begin{indented} \item[] September 2016
\end{indented}

\section{Introduction}


\section{Introduction}

Coulomb force is at the origin of all vital processes in nature.
It is the dominant interaction from the nanometer to the
millimeter scale. Therefore ability to treat quantum Coulombic
systems represents substantial interest for a large community of
scientists. Unfortunately analytical solutions exist only for two
interacting charged particles and accounts for
 only negligible part of the Coulombic systems. Three interacting
charges constitute therefore the simplest Coulombic problem, which
cannot be solved analytically. Solution of a bound state problem
in systems of three charges is well advanced and now reaches
impressive accuracy, which enables to test the tiniest effects
related to relativistic and QCD corrections~\cite{Korobov_PRL116}.
Nevertheless description of the inelastic collisions remains
problematic and yet unresolved by the mathematically rigorous
methods. Inelastic scattering process and particularly three-body
break-up presents unsurmountable challenge. The main complications
arise due to the long-range nature of the Coulomb interaction and
are reflected in: presence of the infinite number of reaction
channels, existence of long-range couplings between the separate
reaction channels (in particular degenerate ones), inability to
construct three-particle break-up asymptotes.

Aforementioned complications naturally emphasize the need in
developing methods of the scattering theory, which does not rely
on a good knowledge of wave function behavior in far asymptotes.
Such methods exist and are in particular successful in describing
collisions dominated by the short-range interactions~\cite{PPNP}.
One of such methods, Complex scaling, is proved to be very
efficient in describing resonance phenomena for atomic systems.
Nevertheless, an originally formulated~\cite{NC69} smooth complex
scaling method is not directly applicable in solving scattering
problems with the long-range interactions. For this purpose
exterior complex scaling method has been
proposed~\cite{ecsm_SIMON1979211} but yet had rather limited
success in solving three-charge particle scattering
problems~\cite{Bartlett_ecsm,Volkov_ecsm}. This is due to the fact
that exterior complex scaling contains several drawbacks from the
formal as well as practical point of view. In particular, exterior
complex method
\begin{itemize}
    \item is limited to a case of central and local interaction
    \item is difficult to use together with the partial-wave expansion
    \item is difficult to generalize for N$\geq$3 particle system
\end{itemize}
Alternatively, the smooth complex scaling method is not concerned
by the aforementioned complications. The goal of this work is to
demonstrate that the smooth complex scaling method can be
successfully employed in describing Coulombic three-body
collisions, thus overcoming its original limitation to the
scattering dominated by the short-range interactions.

\section{The formalism}

\subsection{Faddeev-Merkuriev equations}

In many-particle scattering theory Jacobi coordinates are often
used to simplify mathematical formulation. For a three-body system
three
independent Jacobi coordinate sets $(\vec{x}_{\alpha },\vec{y}%
_{\alpha })$ exist
\begin{eqnarray}
\vec{x}_{\alpha } &=&\sqrt{\frac{2m_{\beta }m_{\gamma }}{(m_{\beta
}+m_{\gamma })m}}(\vec{r}_{\gamma }-\vec{r}_{\beta }); \\
\qquad \overrightarrow{y}_{\alpha } &=&\sqrt{\frac{2m_{\alpha
}(m_{\beta
}+m_{\gamma })}{(m_{\alpha }+m_{\beta }+m_{\gamma })m}}\left[ \vec{r}%
_{\alpha }-{\frac{m_{\beta }\vec{r}_{\beta }+m_{\gamma }\vec{r}_{\gamma }}{%
m_{\beta }+m_{\gamma }}}\right] ,
\end{eqnarray}%
where $m_{\alpha }$ and $\vec{r}_{\alpha }$ are respectively a
mass and position of particle $\alpha$. Particle indexation
$\left( \alpha \beta
\gamma \right) $ represents a cyclical permutation of particle indexes $%
(123) $, whereas a mass-factor $m$ of free choice is introduced
into the former equations in order to retain the  proper unities
for the distances. When studying atomic systems it is convenient
to identify this mass with the mass of an electron $m\equiv
m_{e}$.

In the 60's L.D.~Faddeev formulated a set of
equations~\cite{Faddeev:1960su} to solve the three-particle
scattering problems with short-range potentials. Some twenty years
later original Faddeev equations have been elaborated by
S.P.~Merkuriev~\cite{Me80} to treat long-range interactions.
Merkuriev proposed to split Coulomb potential $V_{\alpha }$ into
two parts (short and long range), $V_{\alpha }=V_{\alpha
}^{s}+V_{\alpha }^{l}$, by means of some cut-off function $\chi
_{\alpha }$.
\begin{equation}
V_{\alpha }^{s}(x_{\alpha },y_{\alpha })=V_{\alpha }^{s}(x_{\alpha
})\chi _{\alpha }(x_{\alpha },y_{\alpha });\qquad V_{\alpha
}^{l}(x_{\alpha },y_{\alpha })=V_{\alpha }(x_{\alpha })[1-\chi
_{\alpha }(x_{\alpha },y_{\alpha })].
\end{equation}%
Using the last identity a set of three Faddeev equations is
rewritten:
\begin{equation}
(E-H_{0}-V_{\alpha }-W_{\alpha })\Psi _{\alpha }=V_{\alpha
}^{s}\sum_{\alpha \neq \beta =1}^{3}\Psi _{\beta };\qquad
W_{\alpha }=V_{\beta }^{l}+V_{\gamma }^{l}  \label{MFE}
\end{equation}%
Here $E$ is a center of mass energy and $H_{0}$ is a free
Hamiltonian of a three-particle system.  In these equations the term $%
W_{\alpha }$ represents a non-trivial long-range three-body
potential. This term includes residual long-range interaction
between the projectile particle $\alpha $ and the target composed
of particles $\left( \beta \gamma \right)$. In order to obtain a
set of equations with compact kernels and which efficiently
separate
wave function asymptotes of different binary particle channels, the function $%
\chi_{\alpha }$ should satisfy certain conditions~\cite{Me80}. To
satisfy these conditions Merkuriev proposed a cut-off function in
a form:
\begin{equation}
\chi _{\alpha }(x_{\alpha },y_{\alpha })=\frac{2}{1+\exp \left[ \frac{%
(x_{\alpha }/x_{0})^{\mu }}{1+y_{\alpha }/y_{0}}\right] },
\end{equation}%
with parameters $x_{0},y_{0}$ and $\mu $, which can be chosen to
be different in each channel $\alpha$. A constrain $\mu>2$ should
be however respected, while the choice of $x_{0}$ and $y_{0}$
remains arbitrary. From the perspective of physics parameter
$x_{0}$ is associated with the effective size of the 2-body
interaction; it makes therefore sense to associate this parameter
with a size of two-body bound state. On the other hand parameter
$y_{0}$ is associated with a size of three-body region, region
where three-particle overlap is important.

Faddeev-Merkuriev (FM) equations, as formulated in eq.(\ref{MFE}),
project wave function's asymptotes of the $\alpha $-$\left( \beta
\gamma \right)$ particle  channels to component $\Psi _{\alpha }$.
The total systems wave function is
recovered by summing the three FM components $\Psi (\vec{x},\overrightarrow{y%
})=\Psi _{1}(\vec{x},\overrightarrow{y})+\Psi _{2}(\vec{x},\overrightarrow{y}%
)+\Psi _{3}(\vec{x},\overrightarrow{y})$. Similarly, by summing up
three equations eq.(\ref{MFE}), formulated for each component
$\Psi_{\alpha }$, the Schr\"{o}dinger equation is recovered.


In order to solve FM equations numerically, it is convenient to
express
each FM component $\Psi _{a}$ in its proper set of Jacobi coordinates $(\vec{%
x}_{\alpha},\vec{y}_{\alpha})$. Further it is practical to employ
partial waves to describe the angular dependence of these
components:
\begin{equation}
\Psi _{\alpha }(\vec{x}_{\alpha },\vec{y}_{\alpha
})=\sum\limits_{l_{x},l_{y}}\frac{f_{\alpha
,l_{x},l_{y}}^{(LM)}(x_{\alpha },y_{\alpha })}{x_{\alpha
}y_{\alpha }}\left\{ Y_{l_{x}}(\widehat{x}_{\alpha })\otimes
Y_{l_{y}}(\widehat{y}_{\alpha })\right\} _{LM}, \label{eq_PW_exp}
\end{equation}%
here $\vec{l}_{x}$ and $\vec{l}_{y}$ are partial
angular momenta associated with the Jacobi coordinates $\vec{x}%
_{\alpha }$ and $\vec{y}_{\alpha }$ respectively. Naturally, total
angular momentum $\vec{L}=\vec{l}_{x}+\vec{l}_{y}$ of the system
should be
 conserved.

Let select out an initial scattering state $\widetilde{\Psi
}_{a}^{(in)}$,  proper to the Jacobi coordinate set $\alpha $
(this feat will be expressed by the Kroneker $\delta _{\alpha ,a}$
function). The scattering
state $(a)$ is defined by a particle $\alpha$, which with momentum $%
q_{\alpha }=\frac{m_{e}}{\hbar ^{2}}\sqrt{E-E_{a}}$ impinges on a
bound particle pair $\left( \beta \gamma \right) $. This bound
state is defined by
a proper angular momentum quantum number $l_{x}^{(a)}$ and binding energy $%
E_{a}$. The relative angular momentum quantum number $l_{y}^{(a)}$
should satisfy triangular conditions,  related with the angular
momenta conservation
condition  $\vec{l}_{x}^{(a)}+\vec{l}_{y}^{(a)}=%
\vec{L}$. Then
\begin{equation}
\Psi _{\alpha }^{(a)}(\vec{x}_{\alpha },\vec{y}_{\alpha })=%
\widetilde{\Psi }_{a}^{(in)}(\vec{x}_{\alpha },\vec{y}_{\alpha
})\delta _{\alpha ,a}+\widetilde{\Psi }_{\alpha }^{(a)}(\vec{x}_{\alpha },%
\vec{y}_{\alpha }) \label{eq_inwave_sep}
\end{equation}%
The standard procedure is to consider for $\widetilde{\Psi }%
_{a}^{(in)}(\vec{x}_{\alpha },\vec{y}_{\alpha })$ a free incoming
wave of particle $\alpha$ with respect to a bound pair of
particles $(\alpha,\beta)$. Nevertheless Coulomb field of particle
$\alpha$ easily polarizes and excites the target, resulting into
long-range coupling between different target
configurations~\cite{Gailitis,Hu_Papp_PRL}. As a result,
scattering wave function in its asymptote may converge very slowly
to a free-wave solution . It might be useful to represent
incoming wave function by distorted waves, which describe more
accurately asymptotic solution. It is, the incoming wave may be
generalized to satisfy a 3-body Schr\"{o}dinger equation:
\begin{equation}
(E-H_{0}-V_{\alpha }-\widetilde{W}_{\alpha })\widetilde{\Psi }%
_{a}^{(in)}\equiv 0  \label{Sc_eq_aux_pot}
\end{equation}%
with some auxiliary long-range potential $\widetilde{W}_{\alpha }(\vec{x}%
_{\alpha },\vec{y}_{\alpha })$. This potential is exponentially
bound in $x_{\alpha }$ direction, therefore it does not contribute
to particle recombination process but may couple different target
states. Such an auxiliary potential can be conveniently expressed
by employing a separable expansion:
\begin{equation}
\widetilde{W}_{\alpha }(\vec{x}_{\alpha },\vec{y}_{\alpha
})=\sum_{a,b}\left\vert \varphi _{a,l_{x}}(\vec{x}_{\alpha
})\right\rangle \lambda _{ab}(y_{\alpha })\left\langle \varphi
_{b,l_{x}}(\vec{x}_{\alpha })\right\vert \label{eq_aux_pot}
\end{equation}%
Radial amplitudes representing a distorted incoming wave
$\widetilde{\Psi }_{a}^{(in)}(\vec{x}_{\alpha },\vec{y}_{\alpha
})$ satisfy  standard boundary condition:
\begin{eqnarray}
\widetilde{f}_{\alpha ,l_{x},l_{y}}^{(in,a)}(x_{\alpha },y_{\alpha
}\rightarrow \infty )&=&\varphi _{a,l_{x}}(x_{\alpha })\widehat{j}%
_{l_{y}}(q_{a}y_{\alpha })\delta
_{l_{y},l_{y}^{(a)}}  \\
&+&\sum_{b}\delta _{\alpha
,b}\widetilde{A}_{b,a}(E)\sqrt{\frac{q_{b}}{q_{a}}}\varphi
_{b,l_{x}}(\vec{x}_{\alpha })\exp (iq_{b}y_{\alpha }-il_{y}\pi
/2)\delta _{l_{y},l_{y}^{(b)}} \nonumber,
\end{eqnarray}
where $\widetilde{A}_{b,a}(E)$ is the scattering amplitude due to
the
auxiliary long-range potential $\widetilde{W}_{\alpha }(\vec{x}_{\alpha },%
\vec{y}_{\alpha })$.  Equation~(\ref{Sc_eq_aux_pot}) is easy to
solve numerically; by projecting it on different target states
dependence on $\vec{x}_{\alpha }$ is eliminated, thus leading to a
standard 2-body coupled channel problem.
By solving  eq.(\ref{Sc_eq_aux_pot}) incoming wave $\widetilde{\Psi }%
_{a}^{(in)}(\vec{x}_{\alpha },\vec{y}_{\alpha })$ is obtained
numerically and may be further employed to solve three-body FM
equations. By setting
expressions~(\ref{eq_inwave_sep}-\ref{Sc_eq_aux_pot}) into
original FM equation~(\ref{MFE}), one obtains:
\begin{equation}
(E-H_{0}-V_{\alpha }-W_{\alpha })\widetilde{\Psi }_{\alpha
}^{(a)}=V_{\alpha
}^{s}\sum_{\alpha \neq \beta =1}^{3}\left( \widetilde{\Psi }_{\beta }^{(a)}+%
\widetilde{\Psi }_{\beta }^{(in)}\delta _{\beta ,a}\right) +(W_{\alpha }-%
\widetilde{W}_{\alpha })\widetilde{\Psi }_{a}^{(in)}\delta
_{\alpha ,a} \label{eq_FM_inh}
\end{equation}

The FM amplitude $\widetilde{f}_{\alpha
,l_{x},l_{y}}^{(a)}(x_{\alpha },y_{\alpha }),$ associated with the
component $\widetilde{\Psi }_{\alpha }^{(a)}(\vec{x}_{\alpha
},\vec{y}_{\alpha })$, in the asymptote contains only outgoing
waves. It may contain two-types of outgoing waves: ones
representing binary process where a particle $\alpha $ is
liberated but a pair of particles $\left( \beta \gamma \right) $
remains bound and outgoing waves representing the break-up of the
system into three free particles:
\begin{eqnarray}
\widetilde{f}_{\alpha ,l_{x},l_{y}}^{(a)}(x_{\alpha },y_{\alpha}\rightarrow \infty )&=&\sum_{b}\delta _{\alpha
,b}\overline{A}_{b,a}^{{}}(E)\sqrt{\frac{q_{b}}{q_{a}}}\varphi_{b,l_{x}^{(b)}}(\vec{x}_{\alpha}) \exp(iq_{b}y_{\alpha}-il_{y}^{(b)}\pi /2) \nonumber\\
&+&A_{a,l_{x},l_{y}}(E,\frac{x_{\alpha }}{y_{\alpha }},\sqrt{x_{\alpha }^{2}+y_{\alpha }^{2}})\exp (iQ\sqrt{x_{\alpha }^{2}+y_{\alpha }^{2}})
\end{eqnarray}
The amplitude $\overline{A}_{b,a}(E)$ represents transition between
the distorted binary channels, whereas the amplitude $A_{a,l_{x},l_{y}}(E,\frac{x_{\alpha }%
}{y_{\alpha }},\sqrt{x_{\alpha }^{2}+y_{\alpha }^{2}})$ is set to
describe three-particle break-up process. These amplitudes can be
extracted from the solution $\widetilde{\Psi }_{\alpha }^{(a)}$ of
the FM equations by applying Green's theorem. In this study, we
will concentrate only on scattering
amplitudes related to the rearrangement reactions. The amplitude $\overline{A}%
_{b,a}(E)$ is given:
\begin{eqnarray}
\overline{A}_{b,a}(E) &=&\sqrt{q_{a}q_{b}}\frac{m_{e}}{\hbar
^{2}}\left\{
\left\langle \Psi ^{(a)}\left\vert E-H_{0}\right\vert \widetilde{\Psi }%
_{b}^{(in)}\right\rangle -\left\langle \widetilde{\Psi }_{b}^{(in)}\left%
\vert E-H_{0}\right\vert \Psi ^{(a)}\right\rangle \right\}   \\
&=&\sqrt{q_{a}q_{b}}\frac{m_{e}}{\hbar ^{2}}\left\langle \Psi
^{(a)}\left\vert \sum_{\alpha }\left\{ \left( V_{\alpha }+\widetilde{W}%
_{\alpha }\right) \delta _{\alpha ,b}-V_{\alpha }\right\}
\right\vert \widetilde{\Psi }_{b}^{(in)}\right\rangle
\label{eq_greens_th}
\end{eqnarray}

The total scattering amplitude is given by:
\begin{equation}
A_{b,a}(E)=\overline{A}_{b,a}(E)+\widetilde{A}_{b,a}(E).
\end{equation}
In terms of this full amplitude partial scattering cross section
for a process $b\rightarrow a$ and a partial wave L is defined by:
\begin{equation}
\sigma _{ab}^{L}(E)=\frac{2\pi a_{0}^{2}}{\frac{m_{\alpha
}(m_{\beta
}+m_{\gamma })}{(m_{\alpha }+m_{\beta }+m_{\gamma })m_{e}}q_{a}^{2}}%
(2L+1)\left\vert A_{a,b}(E)\right\vert ^{2}
\end{equation}
One may also define total inelastic cross section for a collision
$(a)$:
\begin{equation}
\sigma _{a,inel}^{L}(E)=\frac{\pi a_{0}^{2}}{2\frac{m_{\alpha
}(m_{\beta
}+m_{\gamma })}{(m_{\alpha }+m_{\beta }+m_{\gamma })m_{e}}q_{a}^{2}}%
(2L+1)\left( 1-\left\vert 1+2iA_{a,a}(E)\right\vert ^{2}\right)
\end{equation}

\subsubsection{Complex scaling}

The next step is to perform the complex scaling (CS)
transformation on the radial parts of the Jacobi coordinates.
Conventional complex scaling is considered here, defined by a
smooth CS transformation:
\begin{equation}
\widehat{S}_{\theta }=e^{i\theta r\frac{\partial }{\partial r}}=e^{i\theta (x%
\frac{\partial }{\partial x}+y\frac{\partial }{\partial y})},
\label{CSO}
\end{equation}%
where parameter $\theta $ is often referred as CS angle. The free
Hamiltonian after CS operation is simply expressed as:
\begin{equation}
H_{0}^{\theta }=\widehat{S}_{\theta }H_{0}\widehat{S}_{\theta
}^{-1}=e^{-2i\theta }H_{0}.
\end{equation}%
An action of CS operator on some radial function $f(x_{\alpha
},y_{\alpha })$ gives:
\begin{equation}
\widehat{S}_{\theta }f(x_{\alpha },y_{\alpha })=f(x_{\alpha
}e^{i\theta },y_{\alpha }e^{i\theta })
\end{equation}
Complex scaling transformation efficiently handles outgoing waves,
by transforming them into exponentially bound functions. On the
other hand incoming waves become exponentially divergent. In
eq.~(\ref{eq_FM_inh}) incoming wave appears only as an
inhomogeneous term premultiplied with a term containing potential.
Thus for the exponentially bound interactions equation kernel
remains compact even after performing complex scaling operation.
Situation is quite different for
the case of Coulomb interaction. In this case the residual interaction term $%
(W_{\alpha }-\widetilde{W}_{\alpha })$ converges only as a power series in $%
1/y_{\alpha }$ and is not able to compensate exponential
divergence of the incoming wave.

Nevertheless, it is expected that a key part of the collision
happens when a projectile gets close to a target, whereas
asymptotic part of the residual interaction plays only a minor
role in the elastic process. Therefore one may try to screen the
residual interaction term at long distances without expecting
sizeable effect on the scattering observables. Furthermore an
impact of the residual term may be minimized by considering
long-range auxiliary potential $\widetilde{W}_{\alpha }(\vec{x}_{\alpha },%
\vec{y}_{\alpha })$, which includes higher order corrections of
the full residual interaction $W_{\alpha }$. This feat will be
explored in this manuscript.

Technically, the complex scaled FM equations are solved to
determine the CS
transformed FM components $\widetilde{\Phi }_{\alpha }^{(a,\theta )}=%
\widehat{S}_{\theta }\widetilde{\Psi }_{\alpha }^{(a)}(\vec{x}_{\alpha },%
\vec{y}_{\alpha })$ and $\widetilde{\Phi }_{\alpha }^{(a,\theta
)}=\widehat{S}_{\theta }\widetilde{\Psi }_{\alpha }^{(a)}$.
When CS distorted incoming waves are necessary $\widetilde{\Phi }_{a}^{(in,\theta )}=\widehat{S}_{\theta }\widetilde{\Psi }%
_{a}^{(in)}$, they can be calculated numerically by solving CS
system of coupled equations corresponding
eq.~(\ref{Sc_eq_aux_pot}).

To keep kernels of the complex-scaled FM equations compact the
term
$\widehat{S}%
_{\theta }(W_{\alpha }-\widetilde{W}_{\alpha })\widehat{S}_{\theta
}^{-1}$ is screened beyond some fixed radius $y_{\alpha },$ using
following expression:
\begin{equation}
\widehat{S}_{\theta }(W_{\alpha }-\widetilde{W}_{\alpha })\widehat{S}%
_{\theta }^{-1}\rightarrow
\begin{array}{lr}
\widehat{S}_{\theta }(W_{\alpha }-\widetilde{W}_{\alpha })\widehat{S}%
_{\theta }^{-1}; & y<y_{cut} \\
\widehat{S}_{\theta }(W_{\alpha }-\widetilde{W}_{\alpha })\widehat{S}%
_{\theta }^{-1}\exp \left[- \left( \frac{y_{\alpha
}-y_{cut}}{y_{sc}}\right)
^{n}\right]; & y>y_{cut}%
\end{array}%
;
\end{equation}%
with $n>1$. In this work the parameter $y_{cut}$ is chosen in a
range 30-35 a.u., whereas $y_{sc}\in (5-6)$ a.u.
Then:
\begin{eqnarray}
\widehat{S}_{\theta }(E-H_{0}-V_{\alpha }-W_{\alpha
})\widehat{S}_{\theta }^{-1}\widetilde{\Phi }_{\alpha }^{(a,\theta
)}&=&V_{\alpha }^{s}\sum_{\alpha
\neq \beta =1}^{3}\left( \widetilde{\Phi }_{\beta }^{(a,\theta )}+\widetilde{%
\Phi }_{a}^{(in,\theta )}\delta _{\beta ,a}\right) \nonumber \\
&+&\widehat{S}_{\theta
}(W_{\alpha }-\widetilde{W}_{\alpha })\widehat{S}_{\theta }^{-1}\widetilde{%
\Phi }_{a}^{(in,\theta )}\delta _{\alpha ,a} \label{eq_CSFM_inh}
\end{eqnarray}

Scattering amplitudes are calculated by modifying integration
contour in~(\ref{eq_greens_th}) along the Complex rotation line,
giving~\cite{Be73,GKO04}:
{\footnotesize
\begin{eqnarray}
\overline{A}_{b,a}(E) &=&\sqrt{q_{a}q_{b}}\frac{m_{e}}{\hbar ^{2}}%
e^{-6i\theta }\int \left[ \widetilde{\Phi }^{(a,\theta )}(\vec{x},%
\vec{y})\right] ^{\ddagger }\sum_{\alpha }\widehat{S}_{\theta
}\left\{ \left( V_{\alpha }+\widetilde{W}_{\alpha }\right) \delta
_{\alpha
,b}-V_{\alpha }\right\} \widehat{S}_{\theta }^{-1}\widetilde{\Phi }%
_{b}^{(in,\theta )}(\vec{x},\vec{y})\cdot d^{3}xd^{3}y \nonumber\\
&&+\sqrt{q_{a}q_{b}}\frac{m_{e}}{\hbar ^{2}}\left\langle \widetilde{\Psi }%
_{b}^{(in)}\left\vert \sum_{\alpha }\left\{ \left( V_{\alpha }+\widetilde{W}%
_{\alpha }\right) \delta _{\alpha ,b}-V_{\alpha }\right\}
\right\vert \widetilde{\Psi }_{b}^{(in)}\right\rangle
\label{eq_CS_amp}
\end{eqnarray}}

Here an expression $\left[ \widetilde{\Phi }^{(a,\theta )}(\vec{x},%
\vec{y})\right] ^{\ddagger }$ represents a biconjugate function of $%
\widetilde{\Phi }^{(a,\theta )}(\vec{x},\vec{y})$. There is no
need to recalculate these biconjugate functions,  they are easily
obtained from the expression of biconjugate partner, via relation:
\begin{equation}
\left[ f(x,y)\left\{ Y_{l_{x}}(\widehat{x}_{\alpha })\otimes Y_{l_{y}}(%
\widehat{y}_{\alpha })\right\} _{LM}\right] ^{\ddagger
}=f(x,y)\left\{ Y_{l_{x}}(\widehat{x}_{\alpha })\otimes
Y_{l_{y}}(\widehat{y}_{\alpha })\right\} _{LM}^{\ast }
\end{equation}

\subsubsection{Numerical solution using Lagrange-mesh technique}

The functions $\widetilde{f}_{\alpha ,l_{x},l_{y}}^{(a,\theta
)}(x_{\alpha },y_{\alpha })$, representing radial dependence of
complex scaled FM components $\widetilde{\Phi }_{\alpha
}^{(a,\theta )}$, are expanded using Lagrange-mesh
basis~\cite{Baye_bible}:
\begin{equation}
\widetilde{f}_{\alpha ,l_{x},l_{y}}^{(a,\theta )}(x_{\alpha
},y_{\alpha
})=\sum_{i_{x}=1}^{N_{x}}\sum_{i_{y}=1}^{N_{y}}C_{\alpha
,i_{x},i_{y}}^{(a,\theta )}\mathcal{F}_{i_{x}}\left( x_{\alpha
}/h_{x_{\alpha }}\right) \mathcal{F}_{i_{y}}\left( y_{\alpha
}/h_{y_{\alpha }}\right) \label{eq_LM_basis}
\end{equation}%
with $C_{\alpha ,i_{x},i_{y}}^{(a,\theta )}$ representing the
complex expansion coefficients to be determined. The $h_{x_{\alpha
}}$ and $h_{y_{\alpha }}$ are scaling parameters for basis
functions defined as
\begin{equation}
\mathcal{F}_{i}(x)=(-1)^{i}c_{i}\frac{x}{x_{i}}\frac{L_{N}(x)}{x-x_{i}}%
e^{-x/2},
\end{equation}%
In this expression $L_{N}(x)$ represents a $N^{th}$ degree
Laguerre polynomial, whereas $x_{i}$ are its zeroes. The
coefficients $c_{i}$ are fixed by imposing basis functions to be
orthonormal, namely:
\begin{equation}
\int_{0}^{\infty }\mathcal{F}_{i}(x)\mathcal{F}_{i^{\prime
}}(x)dx=\delta _{ii^{\prime }}.
\end{equation}

Set of differential equations~(\ref{eq_CSFM_inh}) is transformed
into a linear algebra problem by projecting their angular
dependence on partial wave basis, defined in
eq.~(\ref{eq_PW_exp}). The radial parts are projected  on
Lagrange-mesh basis, defined in eq.~(\ref{eq_LM_basis}). The
coefficients $C_{\alpha ,i_{x},i_{y}}^{(a,\theta )}$, obtained
after solving linear
algebra problem:
\begin{equation}
(H^\theta-E) C_{\alpha ,i_{x},i_{y}}^{(a,\theta )}=b^{(in,\theta)}
\end{equation}
Here $(H^\theta-E)$ represents a part of CS Hamiltonian acting on
wave function's component $\widetilde{\Phi }_{\alpha }^{(a,\theta )}$,
represented by the coefficients $C_{\alpha ,i_{x},i_{y}}^{(a,\theta )}$.
Inhomogeneous term $b^{(in,\theta)}$ is obtained after projecting in eq.~(\ref{eq_CSFM_inh})
terms containing incoming wave term $\widetilde{\Phi }_{a}^{(in,\theta )}$.
are used to approximate complex scaled FM components $%
\widetilde{\Phi }_{\alpha }^{(a,\theta )}$. Finally, the
transformed components $\widetilde{\Phi }_{\alpha }^{(a,\theta )}$
serve to retrieve scattering amplitudes
employing the integral relation~(\ref{eq_CS_amp}). One may refer
to~\cite{These_Rimas_03,Baye_bible} for a more detailed
description of the numerical methods used in this work.

\section{Results}
\subsection{e+Ps(n=1) scattering}

Electron scattering on positronium constitutes probably the
simplest realistic Coulombic three-body system. This system has
been well explored at low energies, below the first positronium
excitation
threshold~\cite{Ward_JPB,PhysRevA.50.1924,PhysRevA.61.032710,PhysRevA.72.062507,PhysRevA.92.032713,Gilmore2004124}.
Above the positronium excitation threshold only calculations based
on close-coupling method are available~\cite{Gilmore2004124},
which if properly parameterized may provide very accurate results
but in general are not constrained to provide an unique physical solution.
\begin{figure}
\begin{center}
\includegraphics[scale=0.28]{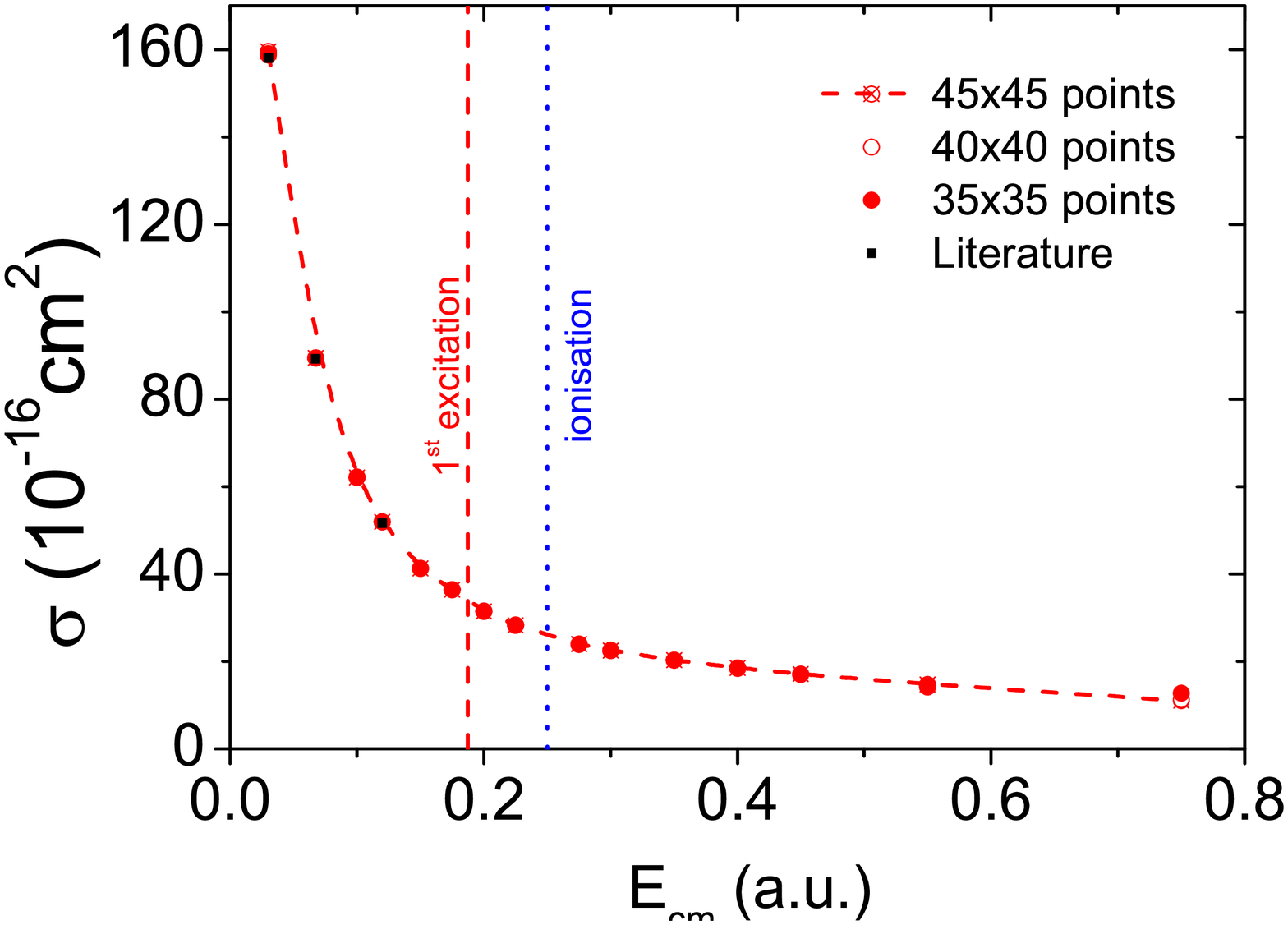}
\includegraphics[scale=0.28]{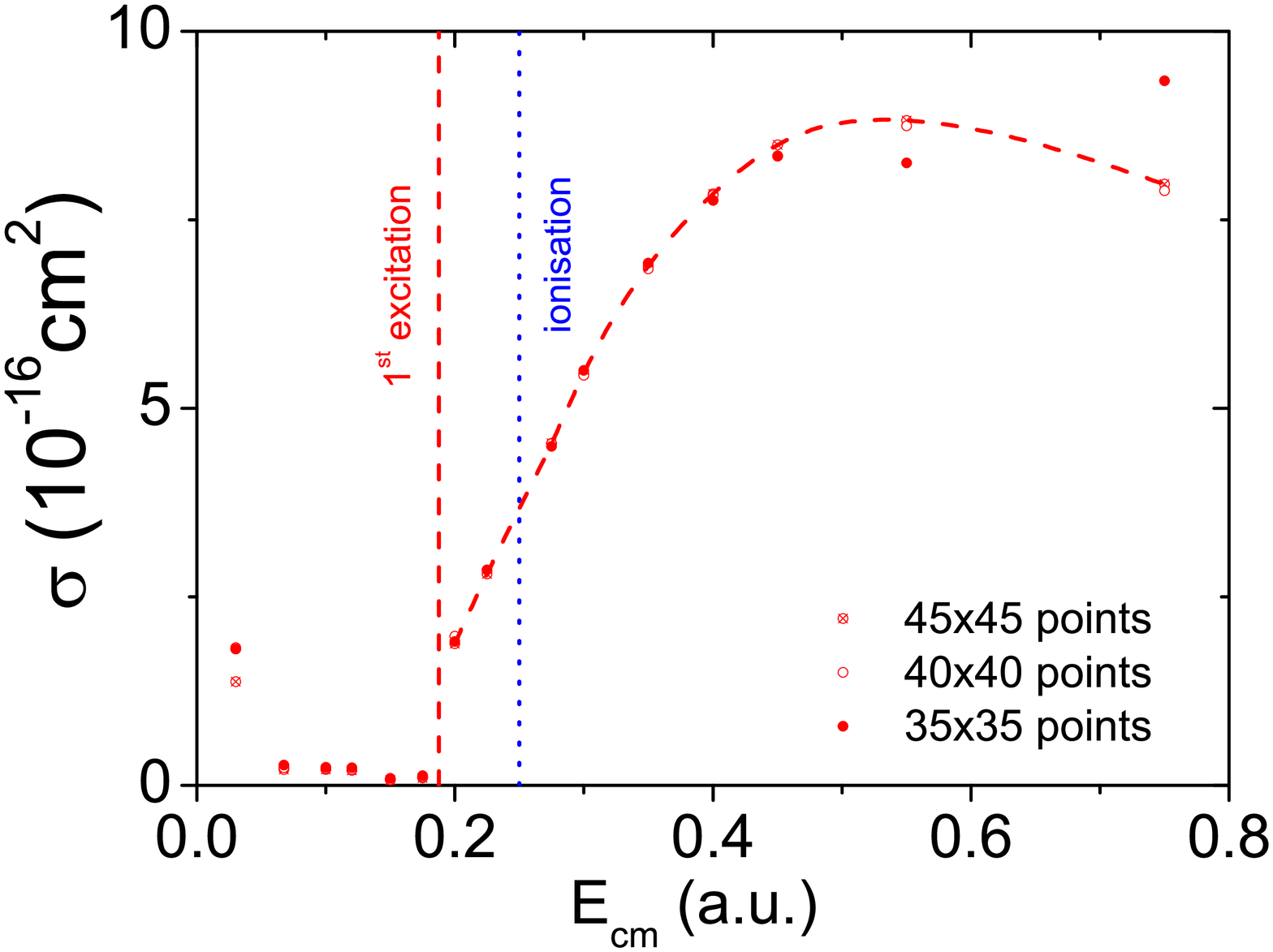}
\caption{\label{fig:eps_cs} Calculated  e+Ps(n=1) total (left panel) and inelastic (right panel)
cross sections in a wide energy range. The dashed lines connecting calculated points is just drawn to guide an eye
and does not represent complete calculation.
 Current calculations are well converged, which becomes clear
by increasing employed Lagrange-mesh basis size. Below Ps(n=2) excitation threshold calculated total cross
sections are compared
with ones compiled from the literature~\cite{Ward_JPB,PhysRevA.61.032710} and represented by full black squares. }
\end{center}
\end{figure}

In figure~\ref{fig:eps_cs}  electron scattering on the ground state of positronium (Ps(n=1)) atom cross section  are presented.
Calculation have been realized
in a broad energy region, starting with a purely elastic case and
spreading well above the positronium ionization threshold.
Below the positronium excitation threshold results of this work
are compared with the most accurate values from literature,
summarized in Table.I of ref.~\cite{PhysRevA.61.032710}.

Present
calculations have been performed by considering free
e+Ps(n=1) waves to represent the incoming wave function in eq.~(\ref{eq_inwave_sep}-\ref{Sc_eq_aux_pot}). The
inhomogeneous term arising from the incoming wave has been
screened in eq.~(\ref{eq_FM_inh}) for e+Ps(n=1) separations beyond $y_eps=35$
a.u. Partial wave expansion has been limited to
max$(l_x,l_y)\leq9$ and proved to be sufficient to get well converged
results, calculations were also limited to total angular momentum $L\leq$5.

 As can be seen in figure~\ref{fig:eps_cs} a basis of 35 Lagrange-Laguerre
 mesh functions in $x$ and
 $y$ direction is enough to get converged results in a broad energy region.
 Only well beyond positronium ionization threshold a basis  35x35 functions
 turns to be insufficient in describing inelastic cross section, nevertheless
 convergence is reached by increasing basis to 40x40 functions.

 As discussed in our previous works employing
 complex scaling~\cite{LC11,PPNP}, large complex scaling angles are not suited
 to perform scattering calculations in $A>2$ particle systems.
 This work reconfirmed this feat. In this work complex scaling parameter
 has been limited to $\theta<10^\circ$, with $\theta=7-8^\circ$
representing an optimal choice. Regardless simplicity of the
 employed approach calculations turn to be very accurate and are in
 line with most accurate published values. The phaseshifts calculated below
 the Ps(n=2) threshold differ from ones reported in~\cite{Ward_JPB,PhysRevA.61.032710} by less than
 0.5$\%$.

 As it is well known, complex scaling operation breaks Hermiticity
 of the Hamiltonian. Consequently the unitarity of S-matrix is not
 provided by the symmetry properties of the CS equations.
 This is the reason why using complex scaling it is more difficult
 to attain numerically the unitarity of  S-matrix  than to
 get highly accurate phaseshifts.  Regardless this fact the
 unitarity of  S-matrix in presented e+Ps(n=1) calculations is
 assured with three-digit accuracy once electron impact energy exceeds
 0.03 a.u. This is clearly demonstrated by
 analyzing inelastic e+Ps(n=1) cross section, extracted
 based on the unitarity property of the S-matrix. In particular,
 inelastic cross section is consistent with a zero value
 in the purely elastic region, below the Ps(n=2) threshold.
 Accurate description of the nearthreshold collisions is naturally the most
 problematic case for the complex-scaling method. After the complex scaling
 operation outgoing waves converge with an exponential factor -$k r sin\theta$, where $k$ is
 a relative momenta of scattering clusters and $r$ is a target-projectile separation distance. This exponent
 vanishes for low impact energies and therefore  approximation of the outgoing waves by
 using square-integrable basis functions becomes inefficient.

\subsection{$e^-$+H scattering}
\begin{figure}
\begin{center}
\includegraphics[scale=0.28]{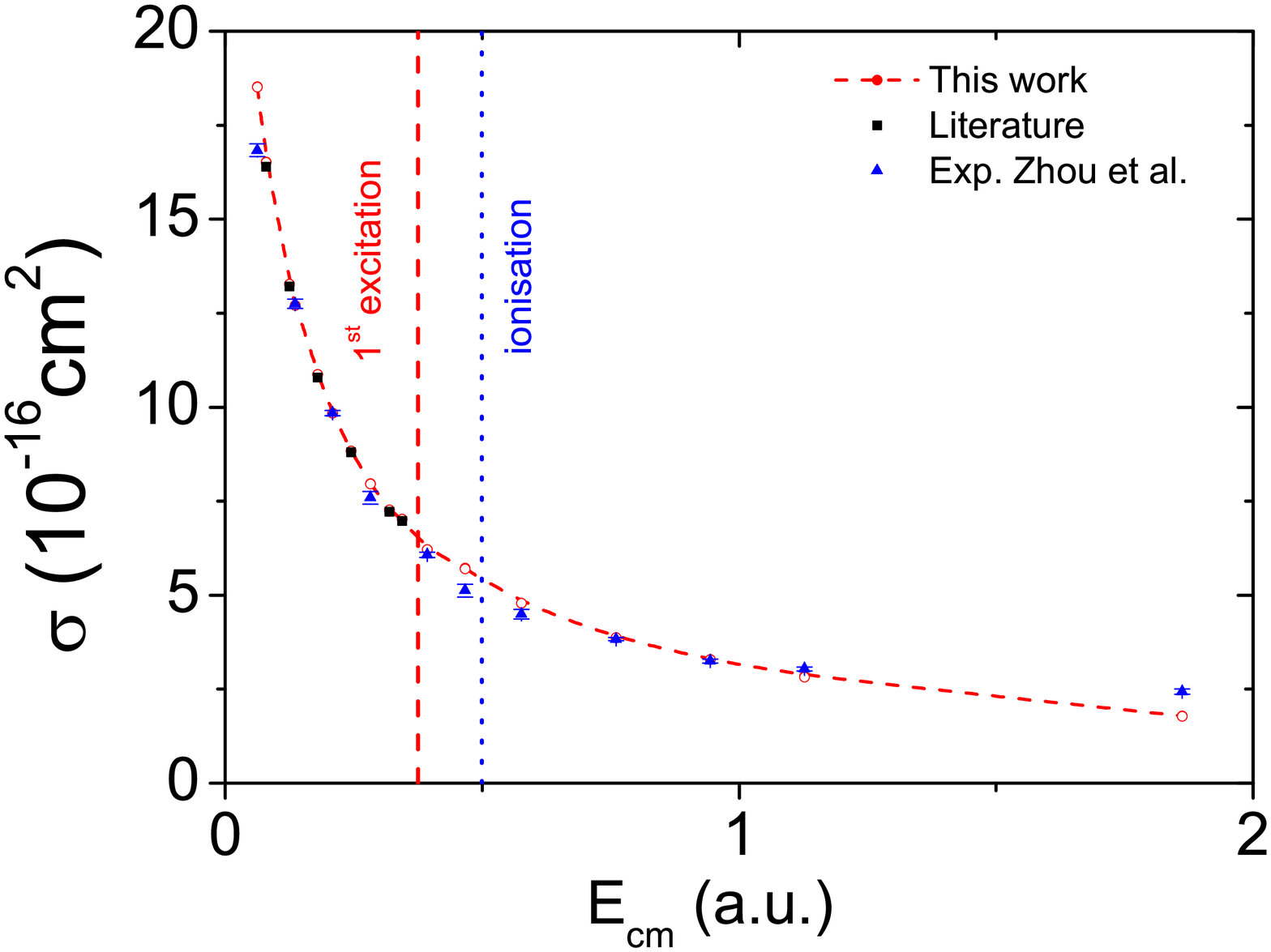}
\includegraphics[scale=0.28]{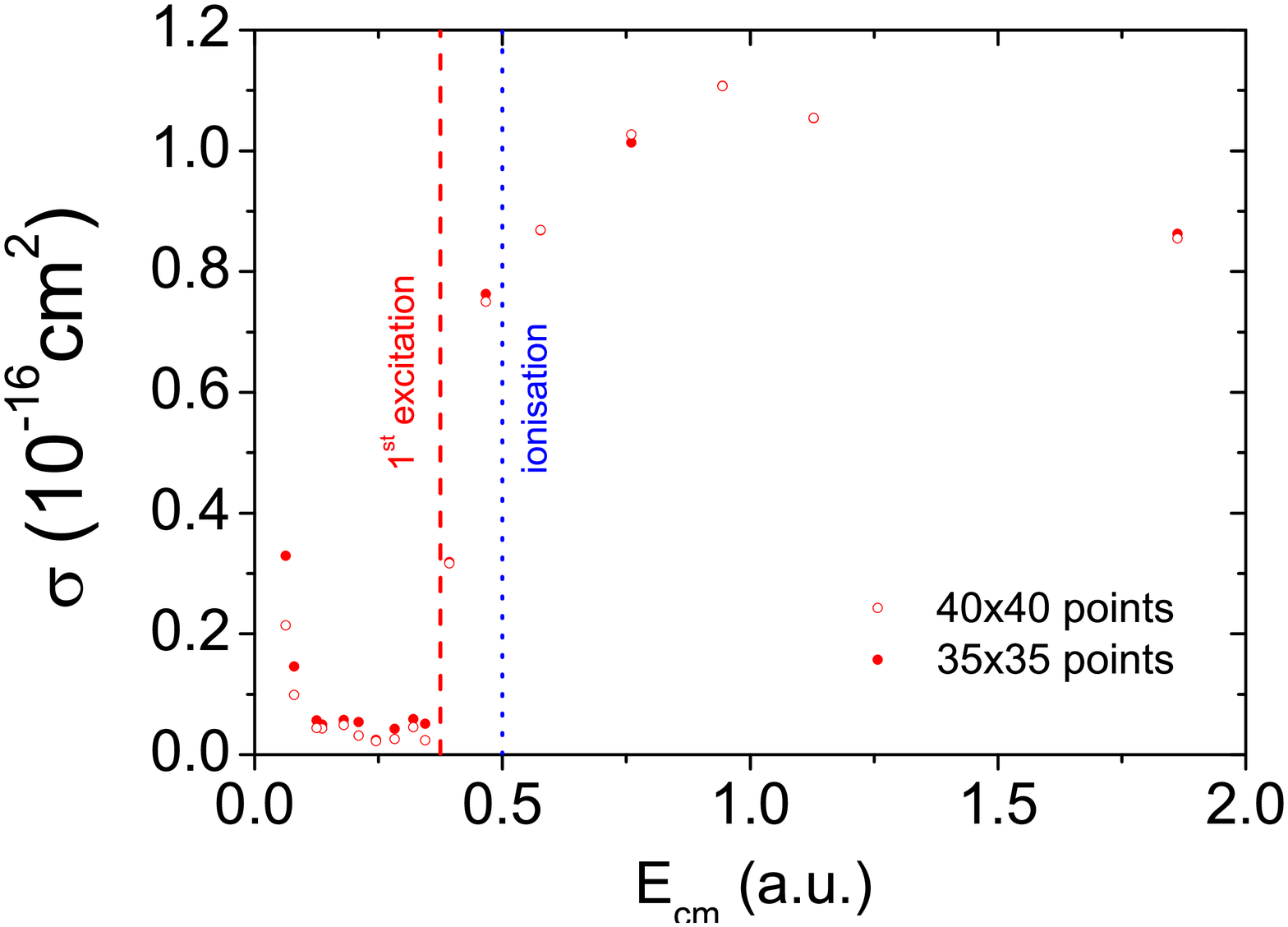}
\caption{\label{fig:eH_cs} The same as in Fig.~\ref{fig:eps_cs} but for  $e^{-}$+H(n=1) scattering.
Calculated values for the total cross section are compared with experimental data of Zhou et al.~\cite{PhysRevA.55.361}.
Below $H(n=2)$ excitation threshold calculated total cross
sections are compared
with ones compiled from the literature~\cite{Gien_JPB} and represented by full black squares. }
\end{center}
\end{figure}
In the figure~\ref{fig:eH_cs} calculations of electron scattering on the ground
sate of Hydrogen atom are presented.
Present calculations have been performed
using the same setup as for e=Ps(n=1) case, described in the last
section. A free e+H(n=1) wave  is considered when separating
inhomogeneous term in eq.~(\ref{eq_inwave_sep}-\ref{Sc_eq_aux_pot}).  35-40
Lagrange-mesh functions were employed for discretizing radial dependence of FM
amplitudes in $x$ and $y$ directions and proved to be enough to get converged results.
Calculated values  agree perfectly with ones found in literature~\cite{Gien_JPB} as well
as with the experimental data of
Zhou et al.~\cite{PhysRevA.55.361}. Only the last point (at 1.87 a.u.) seems to underestimate experimental total cross section.
Total cross section in this energy region, well above the Hydrogen ionization threshold,
has non-negligible contribution of high angular momentum states (beyond $L$=5),
which have not been included in a present calculation.

 The phaseshifts calculated below
 the $H(n=2)$ threshold agree perfectly well with the most accurate calculations
 found in the literature.
 All the phaseshifts fall within the limits defined by the values compiled in reference~\cite{Gien_JPB}.

As pointed out in the previous section, presenting the e+Ps(n=1) scattering, complex scaling technique turns to be
the most difficult to apply at very low energies, close to the threshold. By reducing energy it turns increasingly
difficult to preserve unitarity of the calculated S-matrix. This feat is best demonstrated by
the deviation from the zero-value of the inelastic $e^{-}-H(n=1)$ cross section close to $H(n=1)$ threshold
(see two lowest energy points, situated at $E_cm=0.0624$ and $0.08$ a.u. respectively).
Naturally unitarity of the calculated S-matrix improves once number of basis functions is increased,
nevertheless at very low energies this convergence turns to be rather slow.

\subsection{$e^+$-H(n=1)$\leftrightarrows$p+Ps(n=1) scattering}

\begin{figure}
\begin{center}
\includegraphics[scale=0.28]{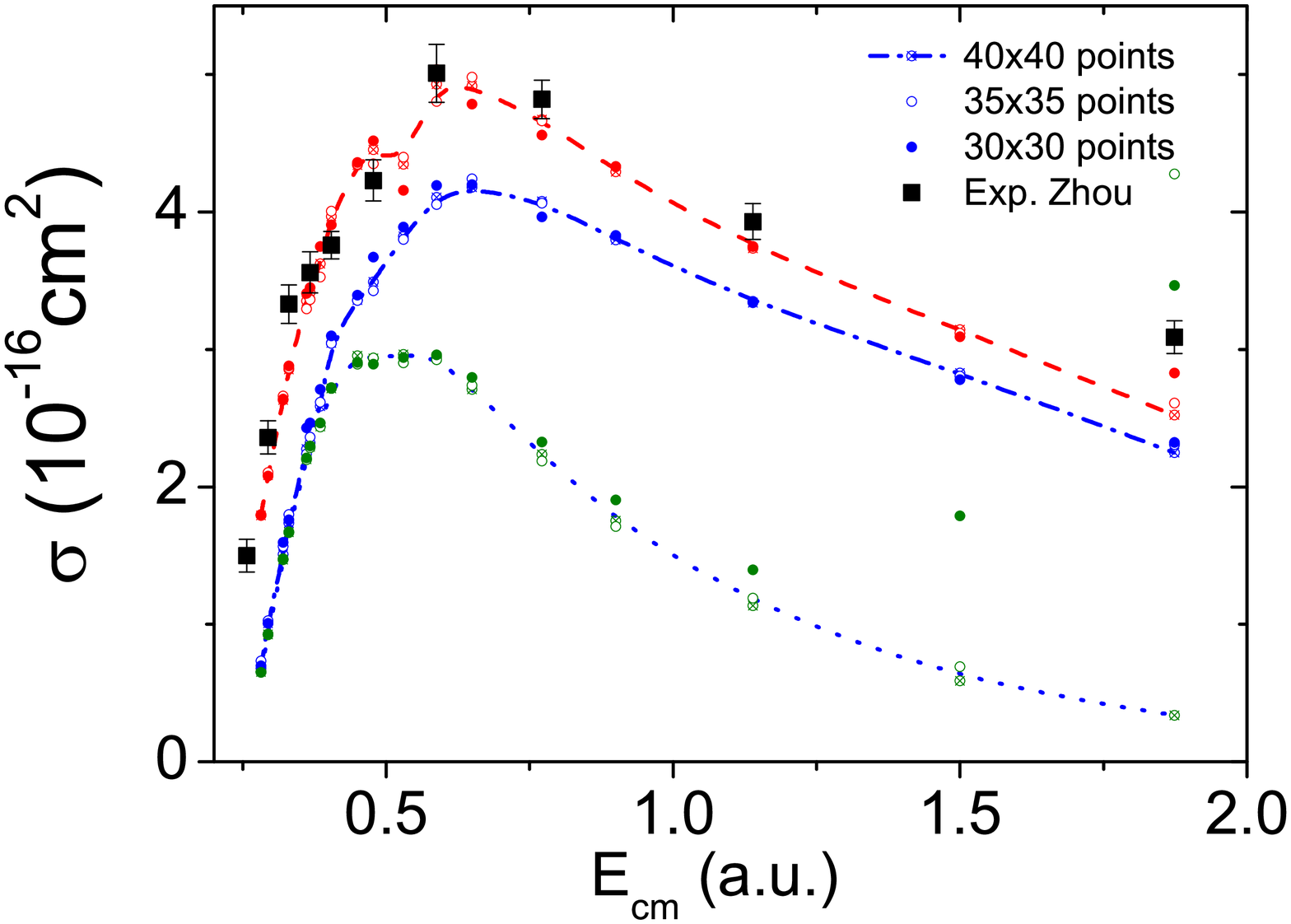}
\includegraphics[scale=0.28]{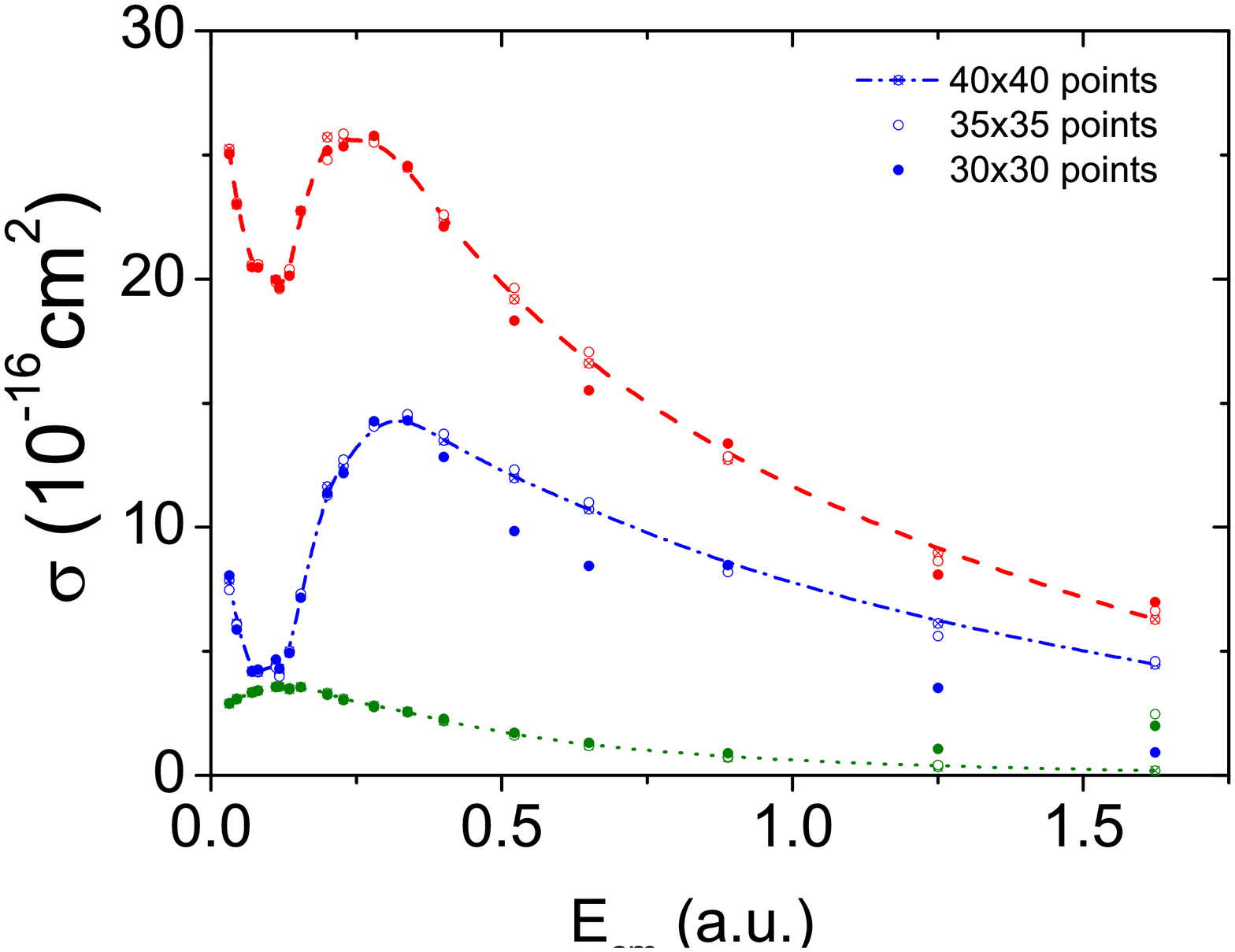}
\caption{\label{fig:pPS_cs}
Study of e$^+$+H(n=1) (left panel) and p+Ps(n=1) (right panel) collisions. Evolution.
of total (red), total inelastic (blue) and  $e^+$+H(n=1)$\leftrightarrows$ p+Ps(n=1) (olive)
cross sections presented with energy. Three sets of calculations performed using
different radial basis sizes.
Calculated total cross section for a e$^+$+H(n=1) process is compared with experimental data of Zhou et al.~\cite{PhysRevA.55.361}. }
\end{center}
\end{figure}

There is an increased interest in studying (anti)proton-positronium collisions in view
of the possible production of antihydrogen atoms. This system was mostly explored
using different variations of close-coupling method~\cite{PhysRevA.93.012709,Mitroy_pPS}; There also exist calculations based on
Hyperspherical-Harmonics~\cite{PhysRevA.50.232},variational
method~\cite{PhysRevA.9.219,0953-4075-30-10-020} as well as Faddeev-Merkuriev equations~\cite{Hu_Papp_PRL,PhysRevA.59.4813}
 but these limited to the energy region of a few lowest energy excitations
of either Hydrogen or Positronium atom.

Elastic $e^+$+H(n=1) collisions below positronium excitation threshold does not present any new features
compared to  $e^-$+H(n=1) or e+Ps(n=1) elastic scattering, discussed in two previous subsections.
Therefore I will concentrate on the energy region above p+Ps(n=1) production threshold.
In  figure~\ref{fig:pPS_cs}  calculations performed by considering only a free  $e^+$+H(n=1) (left panel)
or $p-Ps(n=1)$ (right panel) wave when separating inhomogeneous term in eq.~(\ref{eq_inwave_sep}-\ref{Sc_eq_aux_pot}).

Calculations considering $e^+$+H(n=1) entrance channel are well converged for
a moderate basis of 30x30 Lagrange-mesh functions and does not depend on the variation of
CS parameter in the range $\theta=5-10^\circ$. Results of Present calculations agree perfectly well
with other theoretical calculations as well as with
the experimental data of Zhou et al.~\cite{PhysRevA.55.361}.
Experimental total cross section becomes slightly larger only for the highest energy point calculated,
which has still to non-negligible contribution of high angular momentum scattering not included
in a present calculations. For this point contribution of $L$=7 state, the largest total angular momentum state considered
in this calculation, still accounts for $\approx10\%$ of total cross section,
whereas this state has negligible contribution at lower energies.
Unitarity of the S-matrix
is well preserved, which is demonstrated by the feat that below H(n=2) excitation threshold
inelastic cross section agrees with a Ps(n=1) production one (at these energies Ps(n=1) production
represents the only inelastic channel).
\begin{figure}[h!]
\begin{center}
\includegraphics[scale=0.48]{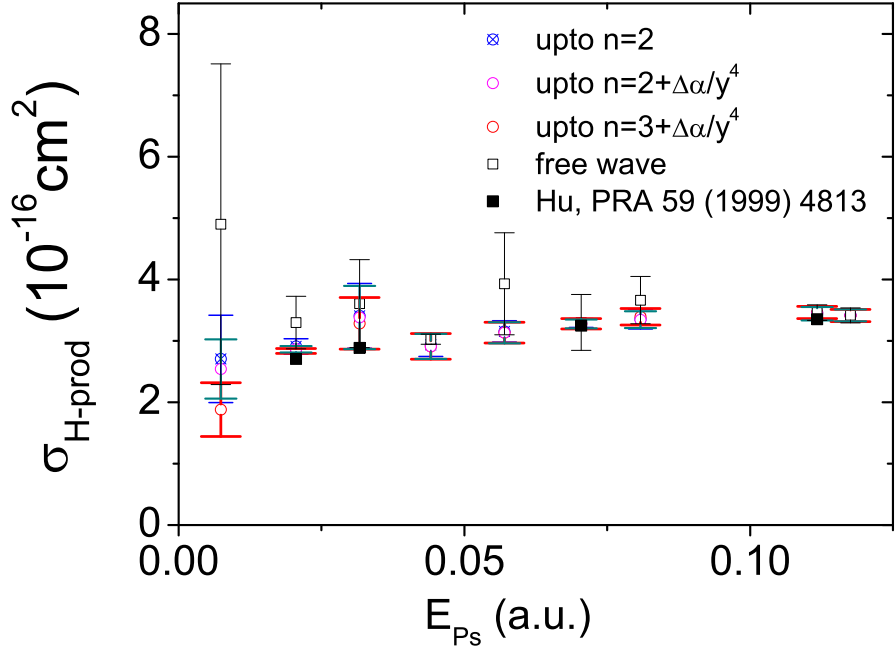}
\end{center}
 \caption{Calculation of the Hydrogen production cross section for p+Ps(n=1) collisions in the Ore gap region.
 Calculations using different assumptions for inhomogeneous term, based on distorted incoming wave of  p+Ps(n=1), were performed.
 These results are compared with calculations of Hu~\cite{PhysRevA.59.4813}, using conventional boundary condition method.}
 \label{Fig_H_prod_err}
\end{figure}

Calculations considering p+Ps(n=1) entrance channel  turns to be less accurate. In particular, problematic are the calculations
performed in the Ore gap region and dominated by relatively low proton(positronium) impact energies.
In this region inelastic p+Ps(n=1) cross section, extracted using unitarity property of S-matrix,
is visibly not converged and improves moderately when increasing
Lagrange-mesh basis size. On the other hand,  Hydrogen production  cross sections calculated from the non-diagonal S-matrix element coupling
$e^+$+H(n=1) and p+Ps(n=1) channels turns to be accurate and well converged even at very low energies.

Even though the low energy region is not the most relevant region to use complex scaling method --
it is still worthy  to pay more attention to the Ore gap region, where p+Ps(n=1) cross sections converge slowly.
In order to improve convergence I have constructed inhomogeneous term in eq.~(\ref{eq_inwave_sep},\ref{Sc_eq_aux_pot}) based on distorted waves
instead of simple free waves used before. Effect of the choice of inhomogeneous term
is studied in figure~\ref{Fig_H_prod_err} by comparing inelastic p+Ps(n=1) cross sections in the problematic  Ore gap region.
These calculations were performed using a basis of
 30x30 Lagrange-mesh functions, with CS parameter set to $\theta=5^\circ$ and total angular momentum
 expansion limited to $L$=3\footnote{This limitation have been used in order to compare the results with ones from ref.~\cite{PhysRevA.59.4813}}.
 Four types of distorted waves, based on the choice of long-range potential in eq.~(\ref{eq_aux_pot}), have been used:
 \begin{itemize}
    \item distorted wave by considering long-range dipole coupling of Ps(n$\leq$2) states,
    with  $\lambda _{ab}(y_{\alpha })=-C_{\alpha }\langle\varphi _{b,l_{x}^{(b)}}(\vec{x}_{\alpha })|\vec{x}_{\alpha }|\varphi _{a,l_{x}^{(a)}}(\vec{x}_{\alpha })\rangle/\tilde{y_{ab}}_{\alpha }^2$\footnote{Expression $\tilde{y_{ab}}_{\alpha }=y_{\alpha }+a_0*n_a^3*n_b^3/y^2_{\alpha }$ has been used to regularize former potential at the origin.
Coefficient $C_{\alpha }$ is a result of the presence of mass scaling factors, present in a definition of Jacobi coordinates $x_{\alpha },y_{\alpha }$.}
    \item considering long-range dipole coupling of  Ps(n$\leq$2) states together with a residual p+Ps(n=1) polarization potential
    \item dipole coupling of Ps(n$\leq$3) states together with a residual p+Ps(n=1) polarization potential
    \item inhomogeneous term based on free wave
 \end{itemize}
 In the figure~\ref{Fig_H_prod_err}  calculated p+Ps(n=1)$\leftrightarrows e^+$+H(n=1) reaction cross section is presented as a range,
  obtained by comparing three different values: cross sections calculated from
 non-diagonal S-matrix elements  ($S_{e^++H(n=1),p+Ps(n=1)}$ and $S_{p-Ps(n=1),e^++H(n=1)}$) as well as cross section
 extracted from diagonal S-matrix element $S_{p-Ps(n=1),p-Ps(n=1)}$ via
 unitarity condition\footnote{In the Ore gap region relation $|S_{p+Ps(n=1),e^++H(n=1)}|^2=1-|S_{p+Ps(n=1),p+Ps(n=1)}|^2$ should hold}.
 One can see that by using distorted waves to construct inhomogeneous term one may considerably improve accuracy of the
 calculated cross sections even at very low
 energies.
 Inclusion of the dipole coupling of Ps(n$\leq$2) states is already enough to get rather accurate results, in agreement with ones from ref.~\cite{PhysRevA.59.4813},
 which were obtained by employing conventional boundary condition approach.
 By considering more complete asymptotic p+Ps(n=1) interaction to determine distorted incoming wave results improve further.

\section{Conclusion}
Coulombic three-body scattering problem have been addressed since the outset of Quantum Mechanics,
however yet have not been fully resolved by mathematically rigorous methods.
In this work it has been demonstrated that conventional smooth complex scaling method can be adapted to solve
Coulombic three-body problems. A novel method has been developed, which combines complex scaling,
distorted wave and Faddeev-Merkuriev equation formalisms. This formalism has been tested
in  studying three realistic Coulombic problems: electron scattering on ground states of Hydrogen
and Positronium atoms as well as a $e^+$+H(n=1) $\leftrightarrows$ p+Ps(n=1)
reaction.

Accurate results were obtained in a wide energy region, also extending
beyond the atom ionization threshold. Calculations for high projectile impact energies
turns to be very accurate and reliable. They agree perfectly with the available experimental
data. On the other hand complex-scaling technique has natural
deficiency in describing very low energy scattering. Still it is demonstrated
that by using distorted incoming waves also description of the scattering process at very low energies
can be significantly improved.

Next challenge is to consider charged particle scattering on the excited states of Hydrogen-type atoms.
 Additional complications should arise due to the presence of long range dipole coupling between
the energy degenerate excited target states. It is not obvious if in such a system inhomogeneous term can be
screened without visible complications as has been done for the scattering on ground state targets in this work.
Distorted wave formalism might be very useful in achieving this goal.

On the other hand smooth complex scaling technique is easily adaptable by any numerical technique and might be easily
incorporated to treat scattering problems in $N\geq3$ systems. Worths noting, that this method
has been  already successfully  adapted to describe the systems
dominated by short-range interactions~\cite{PPNP,PhysRevC.91.041001}. Findings of this work
brings optimism that Coulombic scattering may be finally addressed in the systems
constituting four charged particles. The feat yet never addressed by the ab-initio techniques.

\subsection{Acknowledgments*}
 This  work  was  granted  access  to  the  HPC
resources  of TGCC and IDRIS under  the allocation
2015-x2015056006 made by GENCI. We thank the staff members of the
TGCC and IDRIS for their constant help.


\begin{thebibliography}{10}
\expandafter\ifx\csname url\endcsname\relax
  \def\url#1{\texttt{#1}}\fi
\expandafter\ifx\csname
urlprefix\endcsname\relax\def\urlprefix{URL }\fi
\expandafter\ifx\csname href\endcsname\relax
  \def\href#1#2{#2} \def\path#1{#1}\fi

\bibitem{Korobov_PRL116}
V.~I. Korobov, J.~C.~J. Koelemeij, L.~Hilico, J.-P. Karr,
  \href{http://link.aps.org/doi/10.1103/PhysRevLett.116.053003}{Theoretical
  hyperfine structure of the molecular hydrogen ion at the 1 ppm level}, Phys.
  Rev. Lett. 116 (2016) 053003.
\newblock \href {http://dx.doi.org/10.1103/PhysRevLett.116.053003}
  {\path{doi:10.1103/PhysRevLett.116.053003}}.
\newline\urlprefix\url{http://link.aps.org/doi/10.1103/PhysRevLett.116.053003}

\bibitem{PPNP}
J.~Carbonell, A.~Deltuva, A.~Fonseca, R.~Lazauskas,
  \href{http://www.sciencedirect.com/science/article/pii/S0146641013000938}{Bound
  state techniques to solve the multiparticle scattering problem}, Progress in
  Particle and Nuclear Physics 74 (2014) 55 -- 80.
\newblock \href
  {http://dx.doi.org/http://dx.doi.org/10.1016/j.ppnp.2013.10.003}
  {\path{doi:http://dx.doi.org/10.1016/j.ppnp.2013.10.003}}.
\newline\urlprefix\url{http://www.sciencedirect.com/science/article/pii/S0146641013000938}

\bibitem{NC69}
J.~Nuttal, H.~L. Cohen, Method of complex coordinates for
three-body
  calculations above the breakup threshold, Phys. Rev. 188~(4) (1969)
  1542--1544.
\newblock \href {http://dx.doi.org/10.1103/PhysRev.188.1542}
  {\path{doi:10.1103/PhysRev.188.1542}}.

\bibitem{ecsm_SIMON1979211}
B.~Simon,
  \href{http://www.sciencedirect.com/science/article/pii/0375960179901658}{The
  definition of molecular resonance curves by the method of exterior complex
  scaling}, Physics Letters A 71~(2) (1979) 211 -- 214.
\newblock \href
  {http://dx.doi.org/http://dx.doi.org/10.1016/0375-9601(79)90165-8}
  {\path{doi:http://dx.doi.org/10.1016/0375-9601(79)90165-8}}.
\newline\urlprefix\url{http://www.sciencedirect.com/science/article/pii/0375960179901658}

\bibitem{Bartlett_ecsm}
P.~L. Bartlett,
\href{http://stacks.iop.org/0953-4075/39/i=22/a=R01}{A complete
  numerical approach to electron–hydrogen collisions}, Journal of Physics B:
  Atomic, Molecular and Optical Physics 39~(22) (2006) R379.
\newline\urlprefix\url{http://stacks.iop.org/0953-4075/39/i=22/a=R01}

\bibitem{Volkov_ecsm}
M.~V. Volkov, E.~A. Yarevsky, S.~L. Yakovlev,
  \href{http://stacks.iop.org/0295-5075/110/i=3/a=30006}{Potential splitting
  approach to the three-body coulomb scattering problem}, EPL (Europhysics
  Letters) 110~(3) (2015) 30006.
\newline\urlprefix\url{http://stacks.iop.org/0295-5075/110/i=3/a=30006}

\bibitem{Faddeev:1960su}
L.~D. Faddeev, Scattering theory for a three particle system, Sov.
Phys. JETP
  12 (1961) 1014--1019.

\bibitem{Me80}
S.~P. Merkuriev, Ann. Phys. (N.Y.) 130 (1980) 395.

\bibitem{Gailitis}
M.~Gailitis, R.~Damburg, Sov. Phys. JETP 17 (1963) 1107.

\bibitem{Hu_Papp_PRL}
C.~Y. Hu, D.~Caballero, Z.~Papp,
  \href{http://link.aps.org/doi/10.1103/PhysRevLett.88.063401}{Induced
  long-range dipole-field-enhanced antihydrogen formation in the
  $\overline{p}\phantom{\rule{0ex}{0ex}}+\phantom{\rule{0ex}{0ex}}ps(\mathit{n}\phantom{\rule{0ex}{0ex}}=\phantom{\rule{0ex}{0ex}}2)\phantom{\rule{0ex}{0ex}}\ensuremath{\rightarrow}\phantom{\rule{0ex}{0ex}}{\mathit{e}}^{-}\phantom{\rule{0ex}{0ex}}+\phantom{\rule{0ex}{0ex}}\overline{H}(\mathit{n}\phantom{\rule{0ex}{0ex}}\ensuremath{\le}
  2)$ reaction}, Phys. Rev. Lett. 88 (2002) 063401.
\newblock \href {http://dx.doi.org/10.1103/PhysRevLett.88.063401}
  {\path{doi:10.1103/PhysRevLett.88.063401}}.
\newline\urlprefix\url{http://link.aps.org/doi/10.1103/PhysRevLett.88.063401}

\bibitem{Be73}
T.~Berggren,
  \href{http://www.sciencedirect.com/science/article/pii/037026937390289X}{On
  resonance contributions to sum rules in nuclear physics}, Physics Letters B
  44~(1) (1973) 23 -- 25.
\newblock \href {http://dx.doi.org/10.1016/0370-2693(73)90289-X}
  {\path{doi:10.1016/0370-2693(73)90289-X}}.
\newline\urlprefix\url{http://www.sciencedirect.com/science/article/pii/037026937390289X}

\bibitem{GKO04}
B.~Giraud, K.~Kato, O.~A., J. of Phys. A 37.

\bibitem{Baye_bible}
D.~Baye, The lagrange-mesh method, Physics Reports 565 (2015)
1--107.

\bibitem{These_Rimas_03}
R.~Lazauskas,
  \href{http://tel.ccsd.cnrs.fr/documents/archives0/00/00/41/78/}{Scattering of
  heavy charged particles in atomic and nuclear systems}, Ph.D. thesis,
  Universit\'{e} Joseph Fourier, Grenoble (October 2003).
\newline\urlprefix\url{http://tel.ccsd.cnrs.fr/documents/archives0/00/00/41/78/}

\bibitem{Ward_JPB}
S.~J. Ward, J.~W. Humberstonz, M.~R.~C. McDowell, J. Phys. B 20
(1987) 127.

\bibitem{PhysRevA.50.1924}
C.-Y. Hu, A.~A. Kvitsinsky,
  \href{http://link.aps.org/doi/10.1103/PhysRevA.50.1924}{Resonances in
  ${\mathit{e}}^{\mathrm{-}}$-ps elastic scattering via a direct solution of
  the three-body scattering problem}, Phys. Rev. A 50 (1994) 1924--1926.
\newblock \href {http://dx.doi.org/10.1103/PhysRevA.50.1924}
  {\path{doi:10.1103/PhysRevA.50.1924}}.
\newline\urlprefix\url{http://link.aps.org/doi/10.1103/PhysRevA.50.1924}

\bibitem{PhysRevA.61.032710}
A.~Igarashi, S.~Nakazaki, A.~Ohsaki,
  \href{http://link.aps.org/doi/10.1103/PhysRevA.61.032710}{Phase shifts of
  ${e}^{-}+\mathrm{Ps}$ scatterings and photodetachment cross sections of
  ${\mathrm{ps}}^{-}$}, Phys. Rev. A 61 (2000) 032710.
\newblock \href {http://dx.doi.org/10.1103/PhysRevA.61.032710}
  {\path{doi:10.1103/PhysRevA.61.032710}}.
\newline\urlprefix\url{http://link.aps.org/doi/10.1103/PhysRevA.61.032710}

\bibitem{PhysRevA.72.062507}
A.~Basu, A.~S. Ghosh,
  \href{http://link.aps.org/doi/10.1103/PhysRevA.72.062507}{Doubly excited
  resonant states of positronium negative ion}, Phys. Rev. A 72 (2005) 062507.
\newblock \href {http://dx.doi.org/10.1103/PhysRevA.72.062507}
  {\path{doi:10.1103/PhysRevA.72.062507}}.
\newline\urlprefix\url{http://link.aps.org/doi/10.1103/PhysRevA.72.062507}

\bibitem{PhysRevA.92.032713}
Y.~Zhou, S.~Watanabe, O.~I. Tolstikhin, T.~Morishita,
  \href{http://link.aps.org/doi/10.1103/PhysRevA.92.032713}{Hyperspherical
  calculations of ultralow-energy collisions in coulomb three-body systems},
  Phys. Rev. A 92 (2015) 032713.
\newblock \href {http://dx.doi.org/10.1103/PhysRevA.92.032713}
  {\path{doi:10.1103/PhysRevA.92.032713}}.
\newline\urlprefix\url{http://link.aps.org/doi/10.1103/PhysRevA.92.032713}

\bibitem{Gilmore2004124}
S.~Gilmore, J.~E. Blackwood, H.~Walters,
  \href{http://www.sciencedirect.com/science/article/pii/S0168583X04003787}{Electron/positron
  collisions with positronium}, Nuclear Instruments and Methods in Physics
  Research Section B: Beam Interactions with Materials and Atoms 221 (2004) 124
  -- 128, proceedings of the \{XII\} International Workshop on Positron and
  Positronium Physics.
\newblock \href
  {http://dx.doi.org/http://dx.doi.org/10.1016/j.nimb.2004.03.042}
  {\path{doi:http://dx.doi.org/10.1016/j.nimb.2004.03.042}}.
\newline\urlprefix\url{http://www.sciencedirect.com/science/article/pii/S0168583X04003787}

\bibitem{LC11}
R.~Lazauskas, J.~Carbonell,
  \href{http://link.aps.org/doi/10.1103/PhysRevC.84.034002}{Application of the
  complex-scaling method to few-body scattering}, Phys. Rev. C 84 (2011)
  034002.
\newblock \href {http://dx.doi.org/10.1103/PhysRevC.84.034002}
  {\path{doi:10.1103/PhysRevC.84.034002}}.
\newline\urlprefix\url{http://link.aps.org/doi/10.1103/PhysRevC.84.034002}

\bibitem{PhysRevA.55.361}
S.~Zhou, H.~Li, W.~E. Kauppila, C.~K. Kwan, T.~S. Stein,
  \href{http://link.aps.org/doi/10.1103/PhysRevA.55.361}{Measurements of total
  and positronium formation cross sections for positrons and electrons
  scattered by hydrogen atoms and molecules}, Phys. Rev. A 55 (1997) 361--368.
\newblock \href {http://dx.doi.org/10.1103/PhysRevA.55.361}
  {\path{doi:10.1103/PhysRevA.55.361}}.
\newline\urlprefix\url{http://link.aps.org/doi/10.1103/PhysRevA.55.361}

\bibitem{Gien_JPB}
T.~T. Gien, Observation of a triplet d-wave resonance below the
n=2 h
  excitation threshold in electron–hydrogen scattering, J. Phys. B: At. Mol.
  Opt. Phys. 31 (1998) L629–L635.

\bibitem{PhysRevA.93.012709}
C.~M. Rawlins, A.~S. Kadyrov, A.~T. Stelbovics, I.~Bray,
M.~Charlton,
  \href{http://link.aps.org/doi/10.1103/PhysRevA.93.012709}{Calculation of
  antihydrogen formation via antiproton scattering with excited positronium},
  Phys. Rev. A 93 (2016) 012709.
\newblock \href {http://dx.doi.org/10.1103/PhysRevA.93.012709}
  {\path{doi:10.1103/PhysRevA.93.012709}}.
\newline\urlprefix\url{http://link.aps.org/doi/10.1103/PhysRevA.93.012709}

\bibitem{Mitroy_pPS}
J.~Mitroy, Positronium-proton scattering at low energies, Aust. J.
Phys. 48
  (1995) 893.

\bibitem{PhysRevA.50.232}
A.~Igarashi, N.~Toshima,
  \href{http://link.aps.org/doi/10.1103/PhysRevA.50.232}{Hyperspherical
  coupled-channel study of positronium formation}, Phys. Rev. A 50 (1994)
  232--239.
\newblock \href {http://dx.doi.org/10.1103/PhysRevA.50.232}
  {\path{doi:10.1103/PhysRevA.50.232}}.
\newline\urlprefix\url{http://link.aps.org/doi/10.1103/PhysRevA.50.232}

\bibitem{PhysRevA.9.219}
A.~K. Bhatia, A.~Temkin, H.~Eiserike,
  \href{http://link.aps.org/doi/10.1103/PhysRevA.9.219}{Rigorous precision $p$
  -wave positron-hydrogen scattering calculation}, Phys. Rev. A 9 (1974)
  219--222.
\newblock \href {http://dx.doi.org/10.1103/PhysRevA.9.219}
  {\path{doi:10.1103/PhysRevA.9.219}}.
\newline\urlprefix\url{http://link.aps.org/doi/10.1103/PhysRevA.9.219}

\bibitem{0953-4075-30-10-020}
J.~W. Humberston, P.~V. Reeth, M.~S.~T. Watts, W.~E. Meyerhof,
  \href{http://stacks.iop.org/0953-4075/30/i=10/a=020}{Positron - hydrogen
  scattering in the vicinity of the positronium formation threshold}, Journal
  of Physics B: Atomic, Molecular and Optical Physics 30~(10) (1997) 2477.
\newline\urlprefix\url{http://stacks.iop.org/0953-4075/30/i=10/a=020}

\bibitem{PhysRevA.59.4813}
C.-Y. Hu,
\href{http://link.aps.org/doi/10.1103/PhysRevA.59.4813}{Multichannel
  ${e}^{+}+\mathrm{H}$ calculations via the modified faddeev equations}, Phys.
  Rev. A 59 (1999) 4813--4816.
\newblock \href {http://dx.doi.org/10.1103/PhysRevA.59.4813}
  {\path{doi:10.1103/PhysRevA.59.4813}}.
\newline\urlprefix\url{http://link.aps.org/doi/10.1103/PhysRevA.59.4813}

\bibitem{PhysRevC.91.041001}
R.~Lazauskas,
\href{http://link.aps.org/doi/10.1103/PhysRevC.91.041001}{Modern
  nuclear force predictions for $n-^{3}\mathrm{H}$ scattering above the three-
  and four-nucleon breakup thresholds}, Phys. Rev. C 91 (2015) 041001.
\newblock \href {http://dx.doi.org/10.1103/PhysRevC.91.041001}
  {\path{doi:10.1103/PhysRevC.91.041001}}.
\newline\urlprefix\url{http://link.aps.org/doi/10.1103/PhysRevC.91.041001}

\end{thebibliography}
\end{document}